\documentclass[12pt]{article}

\usepackage{amsmath}
\usepackage{amssymb}
\usepackage[dvips]{graphicx}
\usepackage{rotating}
\usepackage{fancyhdr}

\newcounter{tabl}
\newcommand{\be}{\begin{equation}}
\newcommand{\ee}{\end{equation}}
\newcommand{\beq}{\begin{eqnarray}}
\newcommand{\eeq}{\end{eqnarray}}
\newcommand{\bea}[2]{\be\label{#2}\begin{array}{#1}}
\newcommand{\eea}{\end{array}\ee}

\def\det{\,{\rm det}\, }
\def\diag{{\rm diag}}

\def\({\left(}
\def\){\right)}
\def\[{\left[}
\def\]{\right]}
\def\p{\partial}

\def\11{1\!\! 1}

\def\hf{\frac{1}{2}}

\def\eps{\varepsilon}

   \def\CC {{\cal C}}
   \def\CD {{\cal D}}

   \def\CG {{\cal G}}
   \def\CH {{\cal H}}

   \def\CV {{\cal V}}

\newcommand{\tP}{\lefteqn{\smash{\mathop{\vphantom{<}}\limits^{\;\sim}}}P}

\newcommand{\tN}{\lefteqn{\smash{\mathop{\vphantom{\Bigl(}}\limits_{\sim}
\atop \ }}N}

\newcommand{\tNn}{\lefteqn{\smash{\mathop{\vphantom{\Bigl(}}\limits_{\,\sim}
\atop \ }}{\cal N}}

\newcommand{\nd}{{\cal N}}

\newcommand{\derL}{\Lambda_{(1)}}
\newcommand{\dderL}{\Lambda_{(2)}}
\newcommand{\Cp}{\CC}
\newcommand{\Cpp}{\CC_*}

\begin{document}
%
%

\title{Hamiltonian Analysis of non-chiral Plebanski
Theory and its Generalizations}

\vspace{0.7cm}

\author{Sergei Alexandrov$^1$ and Kirill Krasnov$^2$}

\date{}

\maketitle

\vspace{-1cm}
\begin{center}
$^1$\emph{Laboratoire de Physique Th\'eorique \&
Astroparticules, CNRS UMR 5207,}\\
\emph{Universit\'e Montpellier II, 34095 Montpellier Cedex 05, France}\\
\vspace{0.4cm}
$^2$\emph{Mathematical Sciences, University of Nottingham, Nottingham, NG7 2RD, UK}
\end{center}
\vspace{0.1cm}
\begin{abstract} We consider non-chiral, full Lorentz group-based
Plebanski formulation of general relativity in its version that utilizes
the Lagrange multiplier field $\Phi$ with ``internal'' indices. The Hamiltonian analysis of
this version of the theory turns out to be simpler than in the previously considered in the
literature version with $\Phi$ carrying spacetime indices. We then extend the Hamiltonian analysis to
a more general class of theories whose action contains scalars invariants constructed from $\Phi$.
Such theories have recently been considered in the context of unification of gravity
with other forces. We show that these more general theories have six additional
propagating degrees of freedom as compared to general relativity, something that has not
been appreciated in the literature treating them as being not much different from GR.
\end{abstract}

\

\section{Introduction}

The original Plebanski formulation
of general relativity \cite{Plebanski}, see also \cite{Capovilla:1991qb},
is chiral, i. e., based on the self- anti-self-dual split of the Lie algebra of the Lorentz group.
A formulation using the same key ideas but based on the full Lorentz group has been considered e.g. in
\cite{De Pietri:1998mb}. In this non-chiral, full Lorentz group-based version it was later
generalized in \cite{Freidel:1999rr}
to general relativity in an arbitrary number of spacetime dimensions. This paper also observed that
when one works with the full Lorentz group there are two classically equivalent but distinct
Plebanski-type formulations. Namely, one formulation uses the Lagrange multiplier field with only ``internal''
indices (in the case of 4 spacetime dimensions this is the theory considered in
\cite{De Pietri:1998mb}), while the other formulation uses spacetime indices. It is this latter
version of the full Lorentz group Plebanski formulation that has been mainly considered
as the starting point of the so-called spin foam quantization, due to the fact that its
discretization leads to the so-called simplicity constraints most naturally. The Hamiltonian analysis
of this version of SO(4) Plebanski theory has been carried out in \cite{Buffenoir:2004vx,Alexandrov:2006wt}.

In this short paper we revisit the Hamiltonian analysis of the non-chiral, full Lorentz
group based, Plebanski formulation of general relativity, and perform the analysis
of the version \cite{De Pietri:1998mb} of the theory with internal index Lagrange multiplier
field. The analysis turns out to be simpler than that in \cite{Buffenoir:2004vx},
which is one of our motivations for writing it down.

However, the main purpose of this note is to extend the Hamiltonian treatment to a more
general class of theories, which is as follows. Generalizing the self-dual Plebanski theory,
paper \cite{Krasnov:2006du} by one of the present authors proposed a class of modified gravity theories
with the action including scalar invariants constructed from Plebanski's Lagrange multiplier field.
Paper \cite{Smolin:2007rx} later combined these ideas with the earlier ideas of Peldan
\cite{Peldan:1992iw} on ``unification'' by extension of the internal gauge group. It considered the
following theory based on a gauge group $G$:
\be\label{action-smolin}
S=\int_M \[B^A\wedge F^A - \frac{1}{2} \Phi^{AB} B^A \wedge B^B +
\frac{g}{2} \Phi^{CD} \Phi^{CD} B^A\wedge B^A\],
\ee
where the upper case Latin letters from the beginning of the alphabet are the Lie algebra
ones, $B^A$ is a two-form field, and $F^A$ is the curvature of a $G$-connection
$A^A$, and $\Phi^{AB}$ is the ``Lagrange multiplier'' field that is required to be
traceless $\Phi^{AA}=0$. The author argued that it can be interpreted as a $G/SO(4)$ gauge
theory coupled to gravity. In particular, it was implied that the theory
(\ref{action-smolin}) based on the gauge group $SO(4)$ {\it is} gravity (possibly modified).

The main purpose of this note is to elucidate the nature of this theory for $G$ being the Lorentz group. 
We find, somewhat
surprisingly, that, unlike the modified theories \cite{Krasnov:2006du} based on the
self-dual Plebanski gravity, the non-chiral theory (\ref{action-smolin}) contains
many more propagating degrees of freedom as compared to general relativity. It may
still be possible to interpret it as a gravitational theory, but before such an interpretation
can be possible one has to face a very difficult question of why the additional propagating degrees of freedom
predicted by it are not observed. We do not attempt to develop an interpretation for
such a theory (\ref{action-smolin}) in this short note, our main aim being just
to point out that the theory is much farther from Plebanski's version of general relativity
than one might naively expect.

Our analysis applies not just to (\ref{action-smolin}), but to a more
general class of theories of the type proposed in \cite{Krasnov:2006du} that are parametrized by
a single scalar function of the ``Lagrange multiplier'' field $\Phi$. Their action is
given by:
\be\label{action}
S=\int_M \[ B^A\wedge F^A - \frac{1}{2} (\Phi^{AB} - \Lambda(\Phi) g^{AB}) B^A \wedge B^B\],
\ee
where $\Lambda(\Phi)$ is an arbitrary $G$-invariant scalar function of the traceless
``internal'' tensor $\Phi^{AB}$, and $g^{AB}$ is an invariant metric on the Lie algebra
of $G$. The case considered in \cite{Smolin:2007rx} corresponds to $\Lambda(\Phi)=
g \Phi^{AB} \Phi^{AB}$. In this note we shall study the case of the Lorentz group only,
but the case of an arbitrary gauge group can be treated along the same lines. Let us
also note that the case $G$ being the Lorentz group and $\Lambda=const$ is just the non-chiral Plebanski
theory \cite{De Pietri:1998mb} equivalent to general relativity. Thus, the main
result of this note is that non-chiral Plebanski theory $\Lambda=const$ is a very degenerate
member of a much more general class of theories (\ref{action}), with a generic
theory from this class having six more degrees of freedom than $\Lambda=const$ one.

The organization of this note is as follows. In section \ref{sec_Pleb} we perform
the Hamiltonian analysis of non-chiral Plebanski theory in its version using
the Lagrange multiplier field with internal indices. In section \ref{sec_gen}
we repeat the analysis for the class of the Lorentz group-based generalized
theories (\ref{action}) and show that they contain six more propagating degrees
of freedom as compared to the case $\Lambda=const$ that gives general relativity.

Our conventions and notations are as follows. We consider simultaneously two signatures
that are distinguished by the parameter $\sigma=\pm 1$: it is positive in the
case of the Riemannian signature when the gauge group is $G=SO(4)$ and negative in
the Lorentzian case when $G=SO(3,1)$. In both cases we obtain similar results.
In particular, all the results about
the structure of the phase space of the theory and the number of propagating degrees
of freedom do not depend on the signature.
We use Greek letters for spacetime indices, small Latin letters
from the middle of the alphabet for spatial indices, capital Latin letters from the middle of
the alphabet for internal vector indices,
$I,J,\cdots \in \{1,2,3,4\}$, and small Latin letters from the beginning
of the alphabet as $so(3)$ indices, $a,b,\cdots \in \{1,2,3\}$.
Symmetrization and anti-symmetrization of indices is denoted by $(\cdot\, \cdot)$ and $[\cdot\,\cdot]$ 
correspondingly and both are defined with the weight $1/2$.
The antisymmetric tensor $\eps^{IJKL}$ is normalized such that $\eps^{1234}=1$ and the internal
indices are lowered and raised with the metric $\eta=\diag(\sigma,1,1,1)$. As a result, one obtains
that $\eps^{IJKL}\eps_{IJKL}=\sigma 4!$. A metric
$g_{IJ,KL}$ on the Lie algebra is defined as
$g_{IJ,KL}=(1/2)(\eta_{IK}\eta_{JL}-\eta_{JK}\eta_{IL})$. The structure constants
of $G$ are denoted by $f^{IJ}_{KL,MN}$.
The density $\tilde{\epsilon}^{\mu\nu\rho\sigma}\in\{0,+1,-1\}$ is defined as usual with
$\tilde{\epsilon}^{\mu\nu\rho\sigma}=1$ if $(x^\mu,x^\nu, x^\rho,x^\sigma)$ is a coordinate system
with positive orientation.

\section{Canonical analysis of Plebanski theory}
\label{sec_Pleb}

We consider the non-chiral Plebanski action for general relativity with
a cosmological constant, in its version due to \cite{De Pietri:1998mb} with
the Lagrange multiplier field with internal indices:
\begin{eqnarray}\label{Plebanskiaction}
S_{\rm Pl}[A,B,\varphi] \; = \;  \frac{1}{2} \int_{\cal M} \!\! d^4x\,\,
\tilde{\epsilon}^{\mu \nu \rho \sigma} \[ g_{IJ,KL} B_{\mu \nu}^{IJ}F_{\rho \sigma}^{KL}
+\hf\(\varphi_{IJKL} +\Lambda \eps_{IJKL}\) B_{\mu\nu}^{IJ}B_{\rho \sigma}^{KL}\] .
\end{eqnarray}
In this expression, $\varphi_{IJKL}=\varphi_{[IJ][KL]}$ is an ``internal'' tensor, playing the
role of a field of Lagrange multipliers, see on this below. For this reason we shall refer to
it as a ``Lagrange multiplier'' field, even in the generalized case considered in the next section,
where its Lagrange multiplier role is lost. The field $\varphi_{IJKL}$ must be
symmetric under the exchange of the pair $[IJ]$ with $[KL]$  and is required to satisfy the
following tracelessness condition:
\be
\eps^{IJKL} \varphi_{IJKL}=0.
\label{tracephi}
\ee
The action \eqref{Plebanskiaction} represents general relativity with cosmological constant $\Lambda$
as the topological $BF$ theory with additional constraints (the ``simplicity'' constraints generated
when one varies the action with respect to $\varphi^{IJKL}$) on the 2-form $B$ ensuring that it comes
from a frame field.

The Hamiltonian formulation of this system is obtained as follows.
First, the action is rewritten as
\begin{eqnarray}
\label{firstorderPlebanskiaction}
S'_{\rm Pl}  =
 \int dt \int_{\Sigma} \!\! d^3x \;\(
 \tP^i_{IJ}\partial_t A_i^{IJ} - H\),
\end{eqnarray}
where we have introduced the momentum conjugate to the connection field:
\begin{equation}
\tP^i_{IJ} = g_{IJ,KL}\eps^{ijk}B_{jk}^{KL},
\label{PiB}
\end{equation}
and the canonical Hamiltonian is given by:
\be
-H=A_0^{IJ}D_i\tP^i_{IJ}+B_{0i}^{IJ}\( g_{IJ,KL}\eps^{ijk} F_{jk}^{KL}
+\({\varphi_{IJ}}^{KL}+\Lambda{\eps_{IJ}}^{KL} \)\tP^i_{KL} \),
\label{Hamilt}
\ee
where we have integrated by parts to get the first term.

The main difference between the version of the theory (\ref{Plebanskiaction}) and that analyzed in
\cite{Buffenoir:2004vx} with the Lagrange multiplier field carrying spacetime indices is
that the variables $B_{0i}^{IJ}$ appear in the Hamiltonian form of the action (\ref{firstorderPlebanskiaction})
linearly, while in the other version of the theory they appear quadratically.
Because of this, work \cite{Buffenoir:2004vx} introduces momenta conjugate to  $B_{0i}^{IJ}$,
which complicates the analysis. There is no need for this complication in the case
analyzed here.

Thus, the variables $A_0^{IJ},\ B_{0i}^{IJ}$ and $\varphi_{IJKL}$ have vanishing
conjugate momenta and therefore play the role of Lagrange multipliers.
The variation with respect to these variables generates the following conditions
\beq
\CG_{IJ} &\equiv & D_i\tP^i_{IJ}=\p_i\tP^i_{IJ}+f_{IJ,KL}^{MN} A_i^{KL}\tP^i_{MN} \approx 0,
\label{conG}
\\
C^i_{IJ} & \equiv & g_{IJ,KL}\eps^{ijk} F_{jk}^{KL}
+\({\varphi_{IJ}}^{KL}+\Lambda{\eps_{IJ}}^{KL} \)\tP^i_{KL} \approx 0,
\label{conC}
\\
\Phi^{IJKL} & \equiv & B_{0i}^{(IJ}\tP^{i,KL)}-\frac{\sigma}{4}\,\CV\,\eps^{IJKL}\approx 0,
\label{conPhi}
\eeq
where we used the following definition of the 4-dimensional volume
\be
\CV= \frac{1}{24}\,\eps^{\mu \nu \rho \sigma} \eps_{IJKL}\ B_{\mu \nu}^{IJ}B_{\rho \sigma}^{KL}
=\frac{1}{6}\,{\eps_{IJ}}^{KL} B_{0j}^{IJ}\tP^j_{KL}.
\ee
In the following we shall assume that the volume is non-vanishing.

Now, as usual, the constraints (\ref{conG}) are just generators of the internal gauge rotations.
To disentangle the structure of the other constraints, we note that some of them
can be interpreted as equations fixing the Lagrange multipliers. To see this,
let us first concentrate on the conditions \eqref{conC} and split them into several components.
A convenient way to do this is to note that the quantities $\tP^i_{IJ}$ and $B_{0i}^{IJ}$
form an independent basis in the Lie algebra. This follows from the condition \eqref{conPhi} and
our assumption that the volume $\CV$ is non-zero. Therefore we can use $\tP^i_{IJ}$ and $B_{0i}^{IJ}$
to trade the Lie algebra indices $[IJ]$ for the 3d space indices $i,j$. We start by considering the
following combinations:
\beq
\CH_i &\equiv& -\hf\,g^{IJ,KL}\eps_{ijk}\tP^j_{IJ}C^k_{KL}=-\tP^j_{IJ} F_{ij}^{IJ}\approx 0.
\label{conDi}
\\
\CH_0 &\equiv& B_{0i}^{IJ}C^i_{IJ}=g_{IJ,KL}\eps^{ijk} B_{0i}^{IJ}F_{jk}^{KL}
+6\Lambda\CV+\varphi_{IJKL}\Phi^{IJKL}\approx 0.
\label{conD0}
\eeq
Note that these are just the anti-symmetric part of the projection onto $\tP^i_{IJ}$ and the
trace part of the projection onto $B_{0i}^{IJ}$.
These combinations do not depend on the Lagrange multipliers
(besides the last term that can be weakly dropped) and thus
generate primary constraints. The other independent combinations are the
symmetric part of the projection onto $\tP^i_{IJ}$ and the trace-free part
of the projection onto $B_{0i}^{IJ}$:
\beq
g^{IJ,KL}\tP^{(i}_{IJ}C^{j)}_{KL} = \eps^{(ikl}\tP^{j)}_{IJ} F_{kl}^{IJ}
+\(\varphi^{IJKL}+\Lambda\eps^{IJKL}\)\tP^i_{IJ}\tP^j_{KL}\approx 0,
\\
B_{0i}^{IJ}C^j_{IJ}-\frac{1}{3}\,\delta^j_i  B_{0k}^{IJ}C^k_{IJ} =
g_{IJ,KL}\eps^{lmn} B_{0k}^{IJ}F_{mn}^{KL}\(\delta_i^k\delta^j_l-\frac{1}{3}\, \delta_i^j\delta^k_l\)
\nonumber
\\
\qquad
+\({\varphi_{IJ}}^{KL}+\Lambda{\eps_{IJ}}^{KL} \)\(B_{0i}^{IJ}\tP^j_{KL}-\frac{1}{3}\,
\delta_i^jB_{0k}^{IJ}\tP^k_{KL}\)
\approx 0.
\eeq
Here we have in total $6+8=14$ equations that can be interpreted as those for components of
the Lagrange multiplier field $\varphi_{IJKL}$, of which there is 20. Indeed, the equations allow us to
find the $(\tilde{P}^i \varphi \tilde{P}^j)$ components, of which there is 6, as well as the traceless part
of the $(\tilde{P}^i \varphi B_{0j})$ components, of which there is 8. The remaining 6 components of
the Lagrange multiplier field $\varphi$ are those corresponding to contractions
$(B_{0i}\varphi B_{0j})$.

Next we turn to the conditions \eqref{conPhi}. To deal with these equations, it will be convenient
to introduce the following notations
\be
h\, h^{ij}=\frac{\sigma}{2} \, g^{IJ,KL}\tP^i_{IJ}\tP^j_{KL},
\qquad
h=\det h_{ij},
\label{indmetr}
\ee
\be
\nd^i=-\frac{\sigma}{2h}\,\eps^{ijk}h_{jl}B_{0k}^{IJ}\tP_{IJ}^l,
\qquad
\tN=\frac{\CV}{h}
\label{defin_N}
\ee
and
\be
\Phi^{ij}=-\frac{\sigma}{2}\,\eps^{IJKL}\tP^i_{IJ}\tP^j_{KL}.
\label{realPhi}
\ee

Let us now note that instead of using $\tP^i,B_{0i}$ as the basis in the Lie algebra, we may
as well use the quantities $\tP^i$ together with its Hodge dual. Projecting
the conditions \eqref{conPhi} on $\tP^i$ and its dual, after some simple algebra, we get the following
two equations:
\beq
B_{0i}^{IJ}+\frac{\sigma}{2h}\, h_{ij} g^{IJ,KL} \tP^k_{KL} B_{0k}^{MN}\tP^j_{MN}
-\frac{1}{4}\,\tNn h_{ij}\eps^{IJKL}\tP^j_{KL} &= & 0,
\label{eqPhi1}
\\
\sigma g_{IJ,KL} B_{0j}^{KL}\Phi^{ij}-\hf\, \tP^j_{IJ}{\eps^{KL}}_{MN}B_{0j}^{KL}\tP^i_{MN}+\tNn h \tP^i_{IJ}&= & 0.
\label{eqPhi2}
\eeq
Contracting the first of these equations with $\tP^j_{IJ}$, one finds that
\be
h_{(ik} B_{0j)}^{IJ}\tP^k_{IJ}=-\frac{\sigma}{4}\, \tNn h_{ik} h_{jl} \Phi^{kl} .
\ee
Using this and \eqref{defin_N} in \eqref{eqPhi1}, one obtains
\be\label{B0i}
B_{0i}^{IJ}=\frac{1}{4}\,\tNn h_{ij}\eps^{IJKL}\tP^j_{KL}-\hf\,\eps_{ijk}\nd^j \tP^{k,IJ}
+\frac{1}{8h}\, \tNn h_{ik} h_{jl}\tP^{j,IJ}\Phi^{kl}.
\ee
Substituting this result into equation (\ref{eqPhi2}), one finds that it reduces to the condition
independent of $B_{0i}$\footnote{In fact, in the Riemannian case eq. (\ref{eqPhi2}) has
two additional solutions (as can be seen from the equation (\ref{gensimconstr}) for $\derL=0$)
$$
\Phi^{ij}=\pm 2h h^{ij}.
$$
These are equivalent to conditions that $\tP^i_{IJ}$ is (anti-) self-dual.
Thus, these solutions of the simplicity constraints reproduce the (anti-) self-dual sector of Euclidean 
general relativity. It is interesting that these sectors are contained in the 
non-chiral SO(4) Plebanski formulation without any need to introduce the Immirzi parameter
\cite{Immirzi:1996dr}.
}
\be
\Phi^{ij}=0.
\label{simconstr}
\ee
Thus, we have shown that 20 equations \eqref{conPhi} give 6 primary constraints \eqref{simconstr}
and allow to find 14 out of 18 components of $B_{0i}^{IJ}$ via formula \eqref{B0i}. The remaining
components of these Lagrange multipliers are given by lapse $\tNn$ and shift $\nd^i$, which
are left undetermined in \eqref{B0i}.

Substituting the obtained results for the Lagrange multipliers into the Hamiltonian, one obtains
\be
-H=A_0^{IJ}\CG_{IJ}+\CH_0=A_0^{IJ}\CG_{IJ}+\nd^i \CH_i +\tNn \CH +\lambda_{ij}\Phi^{ij},
\label{Hamiltnew}
\ee
where
\be
\CH=\frac{1}{4}\, h_{ij}\,\eps^{ikl} {\eps^{IJ}}_{KL}\tP^j_{IJ}F_{kl}^{KL}+6\Lambda h
\ee
and $\lambda_{ij}$ is some complicated matrix which will not play any role in the following.
The only important for us fact is that it contains 6 remaining undetermined components of
the Lagrange multiplier field $\varphi_{IJKL}$, i.e., those corresponding to the
projections $(B_{0i}\varphi B_{0j})$.

At this point we get exactly the same system as the one obtained in the covariant canonical
formulation of the Hilbert--Palatini action in \cite{SA} (see also \cite{Alexandrov:2006wt}).
This allows the results on the constraint analysis to be borrowed from this work.
One finds that the primary constraints $\CG_{IJ},\ \CH_i, \ \CH$ do not generate
any further conditions, whereas $\Phi^{ij}$ give rise to 6 secondary constraints
\be
\Psi^{ij}=g^{IJ,KL}\eps^{(ikl}h_{km}\tP^m_{IJ} D_l \tP^{j)}_{KL} \approx  0.
\label{secondPsi}
\ee
The condition of the conservation of $\Psi^{ij}$ then
generates a new constraint. Since the covariant derivative in $\Psi^{ij}$
contains the connection, the commutator of the two constraints \eqref{simconstr} and
\eqref{secondPsi} is non-vanishing. As a result, the tertiary constraint gets a contribution
from the last term in the Hamiltonian \eqref{Hamiltnew} proportional to $\lambda_{ij}$.
As we mentioned, the latter contains the remaining unknown components of $\varphi_{IJKL}$.
Thus, the role of the tertiary constraint is simply to fix these last 6 components of
the Lagrange multiplier field.

Due to the non-vanishing commutator,
the constraints $\Phi^{ij}$ and $\Psi^{ij}$ are of second class and thus can be imposed strongly
provided the symplectic structure was replaced by the one given by Dirac bracket.
The other constraints $\CG_{IJ},\ \CH_i, \ \CH$ are of first class.
Their physical meaning is that $\CG_{IJ}$ generates Lorentz gauge transformations, whereas the other
constraints are responsible for the spatial and temporal diffeomorphisms correspondingly.

The arising structure of the phase space is then as follows. The kinematical phase space is
that of pairs $(\tilde{P}^i_{IJ}, A_i^{IJ})$, with the configuration space -- the space
of $G$-connections -- being $3\times 6=18$ dimensional. We have gauge symmetries as
well as diffeomorphisms acting on this space, with the action generated by first class
constraint each of which reduces the dimension of the configuration space by one. This
leaves us with $18-6-4=8$ dimensional configuration space. On top of this, we have
6+6 second class constraints, each of which reduces the dimension of the phase space by one,
thus leaving us with a two-dimensional configuration physical space, which describes
the two propagating degrees of freedom of general relativity.

\section{Canonical analysis of generalized Plebanski theory}
\label{sec_gen}

We now consider a more general class of theories described in the introduction,
where the cosmological constant $\Lambda$ is replaced by a generic function of the Lagrange
multipliers $\varphi_{IJKL}$
\begin{eqnarray}\label{genaction}
S_{\rm gPl}[A,B,\varphi] \; = \;  \frac{1}{2} \int_{\cal M} \!\! d^4x\,
\tilde{\epsilon}^{\mu \nu \rho \sigma} \[ g_{IJ,KL} B_{\mu \nu}^{IJ}F_{\rho \sigma}^{KL}
+\hf\(\varphi_{IJKL} +\Lambda(\varphi) \eps_{IJKL}\) B_{\mu\nu}^{IJ}B_{\rho \sigma}^{KL}\] .
\end{eqnarray}
Now the action depends on the ``Lagrange multiplier'' fields $\varphi^{IJKL}$ non-linearly.
As is standard in this situation, to facilitate the canonical analysis, it is convenient to introduce
the momenta conjugate to these fields. Thus, we add to the action the following terms:
\be
\int_{\cal M} \!\! d^4x \[ \psi^{IJKL}\p_t \varphi_{IJKL} +\lambda_{IJKL}\psi^{IJKL}  \].
\ee
The first term introduces momenta conjugated to $\varphi_{IJKL}$ which makes them dynamical fields.
The second term imposes constraints that the momenta are vanishing, which
returns us to the original action.

Splitting the time and space coordinates brings the action into the form
\begin{eqnarray}
\label{firstordergenaction}
S_{\rm gPl}  =
 \int dt \int_{\Sigma} \!\! d^3x \;\(
 \tP^i_{IJ}\partial_t A_i^{IJ} +\psi^{IJKL}\p_t \varphi_{IJKL}- H\),
\end{eqnarray}
with the canonical Hamiltonian being
\be
-H=A_0^{IJ}D_i\tP^i_{IJ}+B_{0i}^{IJ}\( g_{IJ,KL}\eps^{ijk} F_{jk}^{KL}
+\({\varphi_{IJ}}^{KL}+\Lambda(\varphi){\eps_{IJ}}^{KL} \)\tP^i_{KL} \)+\lambda_{IJKL}\psi^{IJKL}.
\label{genHamilt}
\ee
The variables $A_0^{IJ},\ B_{0i}^{IJ}$ and $\lambda_{IJKL}$ have vanishing
conjugated momenta and therefore play the role of Lagrange multipliers.
A variation with respect to these variables generates the following conditions
\beq
\CG_{IJ} &\equiv & D_i\tP^i_{IJ} \approx 0,
\label{genconG}
\\
C^i_{IJ} & \equiv & g_{IJ,KL}\eps^{ijk} F_{jk}^{KL}
+\({\varphi_{IJ}}^{KL}+\Lambda(\varphi){\eps_{IJ}}^{KL} \)\tP^i_{KL} \approx 0,
\label{genconC}
\\
\psi^{IJKL} &\approx & 0.
\label{genpsi}
\eeq

As before, the conditions \eqref{genconG} do not involve the Lagrange multipliers and thus give
primary constraints. However, they do not yet give generators of gauge transformations for
all the fields, as they do not act on the ``Lagrange multiplier'' fields $\varphi^{IJKL}$. Thus,
it is convenient to shift them by adding a linear combination of the constraints \eqref{genpsi}:
\be
\tilde\CG_{IJ}=\CG_{IJ}-2 f_{IJ,KL}^{MN} \varphi_{MNPQ}\psi^{KLPQ}.
\label{newG}
\ee
This shift amounts in a simple redefinition of the Lagrange multipliers
\be
\tilde\lambda_{IJKL}=\lambda_{IJKL}-2f_{IJ,PQ}^{MN}A_0^{PQ}\varphi_{MNKL}.
\ee
The new constraints \eqref{newG} generate gauge transformations of all the variables, including $\varphi$ and $\psi$.
This is convenient as, since the Hamiltonian is a gauge scalar, $\tilde\CG_{IJ}$ are stable under its
action and no secondary constraints get produced.

Next we turn to the constraints \eqref{genpsi}. Commuting them with the Hamiltonian, we get additional
conditions
\be
\Phi^{IJKL}  \equiv  B_{0i}^{(IJ}\tP^{i,KL)}
-\frac{\CV}{4}\(\sigma\eps^{IJKL}-24\derL^{IJKL}\)\approx 0,
\label{genconPhi}
\ee
where we have introduced
\be
\derL^{IJKL}:=\frac{\p\Lambda(\varphi)}{\p\varphi_{IJKL}}.
\ee
The conditions \eqref{genconPhi} is what replaces \eqref{conPhi} in the case of usual Plebanski
theory.

We can analyze the consequences of \eqref{genconPhi} using the same procedure and the same
notations \eqref{indmetr}-\eqref{realPhi} as in section \ref{sec_Pleb}. Here the first step is to find an
expression for the Lagrange multipliers $B_{0i}^{IJ}$. One finds:
\beq
B_{0i}^{IJ} &=&
\frac{1}{4}\,\tNn h_{ij}\(\eps^{IJKL}-24\sigma\derL^{IJKL}\)\tP^j_{KL}
-\hf\,\eps_{ijk}\nd^j \tP^{k,IJ}
\nonumber \\
&& \qquad\qquad
+\frac{1}{8h}\, \tNn  h_{ik} h_{jl}\tP^{j,IJ}\(\Phi^{kl}+12\tP^k_{KL}\derL^{KLMN}\tP^l_{MN}\).
\eeq
Unlike the case of usual Plebanski theory, this now explicitly depends on the
``Lagrange multipliers'' $\varphi^{IJKL}$. As in the case of the usual Plebanski theory, the
obtained expression for the quantities $B_{0i}^{IJ}$ leaves 4 of them (the lapse and the
shift) undetermined. Thus, to find them we have utilized $18-4=14$ out of 20 constraints
\eqref{genconPhi}, leaving 6 additional constraints whose meaning is to be clarified.

Using the same procedure that led to the simplicity constraints \eqref{simconstr} we now get
the following 6 additional constraints:
\beq
X^{ij} &\equiv&
\Phi^{ij}-\frac{\sigma}{4h^2}\, \Phi^{ik}h_{km}\Phi^{mn}h_{nl}\Phi^{lj}
\label{gensimconstr}
\\
&& +12\(\star\tP^i_{IJ}+\frac{1}{2h}\,\Phi^{ik}h_{kl}\tP^l_{IJ}\)\derL^{IJKL}(\varphi)
\(\star\tP^i_{KL}+\frac{1}{2h}\,\Phi^{jm}h_{mn}\tP^n_{KL}\)
\nonumber \\
&=&
\hf\(\star\tP^i_{IJ}+\frac{1}{2h}\,\Phi^{ik}h_{kl}\tP^l_{IJ}\)
\(\eps^{IJKL}+24\derL^{IJKL}(\varphi)\)
\(\star\tP^i_{KL}+\frac{1}{2h}\,\Phi^{jm}h_{mn}\tP^n_{KL}\)
\approx 0.
\nonumber
\eeq
Unlike the case of usual Plebanski theory analyzed in the previous section,
the constraints $X^{ij}$ now explicitly depend on $\varphi^{IJKL}$. We will
see that it is this fact that eventually results in the theory having
more propagating degrees of freedom.

Applying now the stabilization procedure to $X^{ij}$, one finds further conditions
\beq
Y^{ij}& \equiv &
12 \tilde\lambda_{IJKL}\(\star\tP^i_{MN}+\frac{1}{2h}\,\Phi^{ik}h_{kl}\tP^l_{MN}\)
\dderL^{IJKL,MNPQ}
\(\star\tP^i_{PQ}+\frac{1}{2h}\,\Phi^{jm}h_{mn}\tP^n_{PQ}\)
\nonumber \\
&&
+\{ X^{ij}, C^k_{IJ}\} B_{0k}^{IJ} \approx 0.
\label{genY}
\eeq

Let us leave for the moment these new conditions and turn to the equations \eqref{genconC}.
Assuming that the independent components of $C^i_{IJ}$ are exhausted by contraction with $\tP^j$, the
trace part of $B_{0j}$ and the traceless part of $\star\tP^j$, we split them into 4 parts as follows:
\beq
\CH_i &\equiv& -\hf\,g^{IJ,KL}\eps_{ijk}\tP^j_{IJ}C^k_{KL}=-\tP^j_{IJ} F_{ij}^{IJ}\approx 0.
\label{genconDi}
\\
\CH_0 &\equiv& B_{0i}^{IJ}C^i_{IJ}=g_{IJ,KL}\eps^{ijk} B_{0i}^{IJ}F_{jk}^{KL}
+6\CV\(\Lambda(\varphi)-\varphi_{IJKL}\derL^{IJKL}\)
+\varphi_{IJKL}\Phi^{IJKL}\approx 0,
\label{genconD0}
\\
\Cp^{ij} &\equiv& g^{IJ,KL}\tP^{(i}_{IJ}C^{j)}_{KL} =\eps^{(ikl}\tP^{j)}_{IJ} F_{kl}^{IJ}
+\(\varphi^{IJKL}+\Lambda(\varphi)\eps^{IJKL}\)\tP^i_{IJ}\tP^j_{KL}\approx 0,
\label{genC1}
\\
\Cpp^{ij} &\equiv&  C^i_{IJ}\star\tP^{j,IJ}-\frac{1}{3}\,h^{ij}h_{kl}  C^k_{IJ}\star\tP^{l,IJ}
\nonumber
\\
&=&
\(\delta^i_k\delta^j_l-\frac{1}{3}\, h^{ij}h_{kl}\) \(\eps^{kmn}  F_{mn}^{IJ}\star\tP^l_{IJ}
+\tP^k_{IJ}\varphi^{IJKL}\star \tP^l_{KL}\)
\approx 0.
\label{genC2}
\eeq
All of these conditions are primary constraints. Note that the dependence of the constraint
$\CH_0$ on the fields $\varphi^{IJKL}$ is that of the Legendre transform of the function
$\Lambda(\varphi)$, the phenomenon also observed in the case of self-dual theory in
\cite{Krasnov:2007cq}. The stabilization procedure applied to the last two
constraints produces further conditions which can be written as
\beq\label{genC3}
\tilde\lambda_{IJKL}\(\tP^i_{IJ}\tP^j_{KL}-2\sigma\Phi^{ij}\derL^{IJKL}\)
+\{\Cp^{ij},A_0^{IJ}\tilde\CG_{IJ}+\CD_0  \}&=& 0,
\\ \label{genC4}
\(\delta^i_k\delta^j_l-\frac{1}{3}\, h^{ij}h_{kl}\)\tP^k_{IJ}\tilde\lambda^{IJKL}\star \tP^l_{KL}
+\{\Cpp^{ij},A_0^{IJ}\tilde\CG_{IJ}+\CD_0  \}&=&0,
\eeq
which gives in total $6+8=14$ conditions. For non-vanishing $\dderL$, these conditions together with
6 conditions \eqref{genY} allow one to find all 20 Lagrange multipliers $\lambda_{IJKL}$.
Thus, the secondary constraints \eqref{genY}, \eqref{genC3}, \eqref{genC4} do not contain
constraints on canonical variables and do not generate any further conditions. On the
other hand, the set of constraints $X^{ij}, \Cp^{ij}, \Cpp^{ij}$ allows one to find
all the components of the ``Lagrange multiplier'' field $\varphi^{IJKL}$.
They are of second class because they do not commute with $\psi^{IJKL}$.

All this is
in contrast with what was happening in the case of the usual Plebanski theory, where
the constraints $X^{ij}$ were $\varphi^{IJKL}$ independent, and thus gave constraints
on the phase space variables $\tilde{P}^i_{IJ}$. Their commutator with the Hamiltonian
resulted in secondary second class constraints. And only the condition that those secondary second
class constraints are preserved under the evolution allowed one to determine the remaining
6 components of the Lagrange multiplier field $\varphi^{IJKL}$. In the case of generalized theory we
are now considering, the situation is simpler, in spite of the seeming complexity of all
the equations. Indeed, all the constraints are now simply equations allowing to determine
the Lagrange multipliers $\varphi^{IJKL}$ and $\lambda^{IJKL}$ and do not
generate any constraints on the other phase space variables. In particular,
the stabilization procedure finishes one step earlier than in the case
of the usual Plebanski theory.

It remains to consider the constraints $\CH_i$ and $\CH_0$.
For the first set of constraints $\CD_i$, it is possible to shift them by means of
other constraints in such way that they become generators of spatial diffeomorphisms
and thus stable under the time evolution. For this we define:
\beq
\CD_i &\equiv &  \CH_i+A_i^{IJ} \tilde G_{IJ}-\psi^{IJKL} D_i\varphi_{IJKL}
\\
&=&\CH_i+A_i^{IJ} G_{IJ}-\psi^{IJKL}\p_i\varphi_{IJKL}.
\nonumber
\eeq
The constraint $\CH_0$ is replaced by the full Hamiltonian with the Lagrange multipliers $\varphi_{IJKL}$ fixed
by the previous equations
\be
\CD_0\equiv H(\varphi=\varphi(A,\tilde{P})).
\ee

The structure of the arising phase space is then as follows. Solving all second class constraints
and conditions on the Lagrange multipliers,
one determines all of the components of the fields $\varphi^{IJKL}, \lambda^{IJKL}$.
In addition, the momentum conjugate to $\varphi^{IJKL}$ is zero. The reduced phase space is then
parametrized by pairs $(\tilde{P}^i_{IJ}, A_i^{IJ})$ with the set of first class constraints
$\CG_{IJ},\ \CD_i,\ \CD_0$ acting on it. This is similar to what we have seen in the case of the
usual Plebanski theory, but the key difference now is that there are no additional second
class constraints on the phase space variables. The dimension of the physical configuration
space is then $18-6-4=8$, which is the 2 degrees of freedom available in the usual Plebanski
theory plus additional six propagating DOF.

As we have seen, the question of the number of degrees of freedom described by the theory
(\ref{genaction}) crucially depends on the properties of the matrix of second derivatives
$\Lambda^{IJKL,MNPQ}_{(2)}$ of the function $\Lambda(\varphi)$. Our result about the
number of DOF certainly applies to the case of the quadratic such function considered in
\cite{Smolin:2007rx}, as the matrix of second derivatives in this case is just the identity matrix
in the appropriate space. It would be interesting to know if there are some other choices
of $\Lambda(\varphi)$ (apart from the ``trivial'' constant function) that lead to
theories with two propagating DOF. More generally, it would be interesting to
characterize the ``landscape'' of functions $\Lambda(\varphi)$ in terms of
the number of DOF that the corresponding theory would produce. We leave this
interesting problem to future research. Another interesting problem is to find an interpretation 
of these additional degrees of freedom.

Let us conclude by
reiterating our main message: a general theory from the class (\ref{genaction}) is
very far from the usual Plebanski theory, as it contains many more
propagating DOF. Whether such a more non-trivial theory can be meaningfully
interpreted as a gravity theory only the future can tell.

\section*{Acknowledgements}
We are grateful to Lee Smolin and Simone Speziale for interesting discussions.
The research of the authors is supported in part by CNRS. KK was supported by an EPSRC advanced
fellowship.

\end{document}